\documentclass{emulateapj}
\usepackage{amssymb,amsmath,amsthm}
\usepackage{longtable}
\tabletypesize{\scriptsize}
\usepackage{color}
\usepackage{pdfcolmk}
\definecolor{DarkGreen}{rgb}{0.0, 0.5, 0.0}
\definecolor{purple}{rgb}{0.6, 0.0, 0.4}

\newcommand{\m}{\mbox{$^{\mbox{m}}$}}

\newcommand{\degree}{^\circ}
\newcommand{\rms}{{\it rms}}
\newcommand{\snr}{{\it snr}}

\newcommand{\gsim}{\gtrsim}

\newcommand{\be}{\begin{equation}}
\newcommand{\ee}{\end{equation}}
\newcommand{\sigE}{{\rm $\sigma_8$}}
\newcommand{\msigE}{{\rm \sigma_8}}

\def\sigmaLim{5}
\def\noiseLim{0.75}

\def\mason09{M09}
\def\white03{W03}
\def\holder08{H08}
\def\shaw09{S09}
\def\m10{M10}
\def\c07{C07}

\begin{document}

\title{Cosmological Constraints from a 31~GHz Sky Survey with the Sunyaev-Zel'dovich Array}

\author{
Stephen~Muchovej\altaffilmark{1,2},
Erik~Leitch\altaffilmark{3},
John~E.~Carlstrom\altaffilmark{3,4}, 
Thomas~Culverhouse\altaffilmark{3},
Chris~Greer\altaffilmark{3},
David~Hawkins\altaffilmark{1},
Ryan~Hennessy\altaffilmark{3},
Marshall~Joy\altaffilmark{5}, 
James~Lamb\altaffilmark{1}, 
Michael~Loh\altaffilmark{3},
Daniel~P.~Marrone\altaffilmark{3,6},
Amber~Miller\altaffilmark{7},
Tony~Mroczkowski\altaffilmark{2,8},
Clem~Pryke\altaffilmark{3},
Matthew~Sharp\altaffilmark{3},
David~Woody\altaffilmark{1}
}
\altaffiltext{1}{California Institute of Technology, Owens Valley Radio Observatory, Big Pine, CA 93513}
\altaffiltext{2}{Department of Astronomy, Columbia University, New York, NY 10027}
\altaffiltext{3}{Department of Astronomy \& Astrophysics, Kavli Institute for Cosmological Physics, University of Chicago, Chicago, IL 60637}
\altaffiltext{4}{Department of Physics, Enrico Fermi Institute, University of Chicago, Chicago IL 60637}
\altaffiltext{5}{Space Sciences - VP62, NASA Marshall Space Flight Center, Huntsville, AL 35812}
\altaffiltext{6}{Hubble Fellow}
\altaffiltext{7}{Columbia Astrophysics Lab, Department of Physics, Columbia University, New York, NY 10027}
\altaffiltext{8}{Department of Physics \& Astronomy, U Penn, Philadelphia, PA 19104}

\begin{abstract}
We present the results of a ~6.1 square degree survey for clusters of
galaxies via their Sunyaev-Zel'dovich (SZ) effect at 31~GHz.  From
late 2005 to mid 2007 the Sunyaev-Zel'dovich Array (SZA) observed four
fields of roughly 1.5 square degrees each.  One of the fields shows
evidence for significant diffuse Galactic emission, and we therefore
restrict our analysis to the remaining 4.4 square degrees.  We
estimate the cluster detectability for the survey using mock
observations of simulations of clusters of galaxies; and determine
that, at intermediate redshifts ($z\sim$0.8), the survey is 50\%
complete to a limiting mass (${\rm M_{200 \overline{\rho}}}$) of ${\rm
\sim 6.0\times 10^{14}} M_{\odot}$, with the mass limit decreasing at
higher redshifts.  We detect no clusters at a significance greater
than 5 times the \rms\ noise level in the maps, and place an upper
limit on \sigE, the amplitude of mass density fluctuations on a scale
of 8$h^{-1}$ Mpc, of ${\rm 0.84 + 0.07}$ at 95\% confidence, where the
uncertainty reflects calibration and systematic effects.  This result
is consistent with estimates from other cluster surveys and CMB
anisotropy experiments.

\end{abstract}


\section{Introduction}
\label{sec:intro}

The number density of massive clusters of galaxies depends strongly on
cosmology, in particular through the normalization of the matter power
spectrum, and through the dependence of the volume element on the
geometry of the universe.  As the low-energy photons in the CMB
traverse the hot (${\rm \sim 10^8~K}$) gas of a massive cluster, about
1\% of the photons are inverse-Compton scattered.  The result is a
distortion in the CMB spectrum, the magnitude of which is proportional
to the integrated pressure of the intra-cluster medium (ICM), i.e.,
the density of electrons along the line of the sight, weighted by the
electron temperature (\cite{sunyaev72,sunyaev80}; see also
\cite{birkinshaw1999}).  The SZ flux of a cluster is therefore a
measure of its total thermal energy.

The change in the observed brightness of the CMB due to the SZ
effect is given by
\begin{equation}
\label{eq:y}
\frac{\Delta T_{\rm CMB}}{T_{\rm CMB}}= f(x)\int \sigma_{\rm T}n_e\frac{k_BT_e}{m_ec^2} dl \equiv f(x)y
\end{equation}
where $T_{\rm CMB}$ is the cosmic microwave background temperature,
$\sigma_{\rm T}$ is the Thomson scattering cross section, $k_B$ is
Boltzmann's constant, $c$ is the speed of light, $m_e$, $n_e$, 
$T_e$ are the electron mass, number density and temperature, respectively, and
$f(x)$ contains the frequency dependence of the SZ effect (where ${\rm x \equiv \frac{h\nu}{k_B T_{CMB}}}$).
Equation \ref{eq:y} defines the Compton $y$-parameter.  The SZ
effect appears as a temperature decrement at frequencies below
$\approx 218~{\rm GHz}$, and as an increment at higher frequencies.

As seen in Equation~\ref{eq:y}, the ratio of $\Delta T/T$ is
independent of the distance to the cluster.  This means that the SZ
effect is redshift-independent in both brightness and frequency,
offering enormous potential for finding high-redshift clusters. A
cluster catalog resulting from an SZ survey of uniform sensitivity has
a cluster mass threshold that is only weakly dependent on redshift for
$z\gsim 0.7$ (via the angular diameter distance).  As a result, SZ
cluster surveys are approximately mass-limited and therefore
potentially powerful probes of cosmology
\cite[e.g.,][]{carlstrom2002}.

Experiments such as the South Pole Telescope \citep{ruhl04} and
the Atacama Cosmology Telescope \citep{fowler2004} are surveying
hundreds of square degrees of sky searching for galaxy clusters through their SZ effect.  A
%precursor to these experiments was one performed with the
precursor survey to those being performed by SPT and ACT was performed with the
Sunyaev-Zel'dovich Array (SZA), an 8-telescope interferometer designed
specifically for detecting the SZ effect towards clusters of galaxies.
Over the span of two years, the SZA surveyed a small region of
sky at 31~GHz.  This survey has been valuable in characterizing both
compact and diffuse cm-wave CMB foregrounds.  It has resulted in a
measurement of the power spectrum of the CMB at small scales
\citep{sharp2010}, a characterization of extragalactic compact source
populations \citep{muchovej2010}, and further evidence for large-scale
dust-correlated microwave emission (Leitch et al.  2010, in prep).  In
this paper, we present results of the survey as they pertain to
cosmological parameter estimation.

The paper is organized as follows: in \S 2 we describe the SZA
observations, including a brief description of the instrument, data
reduction and calibration, data quality tests, and foreground source
extraction.  We present the results of the survey in \S 3, followed by
the calculation of the expected number of clusters in \S 4.  In \S 5
we determine a constraint on $\sigma_8$ and systematic uncertainties
associated with this analysis are presented in \S 6.  We discuss our
results in \S 7.

\section{Observations}

\label{sec:surveyBack}
\subsection{The Sunyaev-Zel'dovich Array}
The Sunyaev-Zel'dovich Array (SZA) is an eight-element interferometer
located at Caltech's Owens Valley Radio Observatory near Big Pine,
California.  The observations presented in this analysis were obtained
with wide-band receivers operating from 27--35~GHz.  These receivers
employ high-electron mobility transistor (HEMT) amplifiers
\citep{pospieszalski1995}, with characteristic receiver temperatures
$T_{rx} \sim11-20~{\rm K}$.  Typical system temperatures, including
noise due to the atmosphere, are on order 40-50~K.  
Six of the eight antennas are arranged in a close-packed
configuration (spacings of 4.5-11.5~m), yielding 15 baselines
sensitive to arcminute scales.  Two outer antennas yield baselines of
up to 65~m, for simultaneous detection of contaminating compact
sources at a resolution of $\sim 22.5\arcsec$ (see
\cite{muchovej2007} for details of the array configuration).

This hybrid configuration was chosen to optimize surveying speed, and
can be thought of as a superposition of two interferometers: the
close-packed array acts as a filter for objects of arcminute extent on
the sky, such as clusters of galaxies, and are used in generating {\it
short baseline maps}; the longer baselines provide high-resolution
sensitivity to compact sources, and provide the data for the {\it long
baseline maps}.  The resulting maps from the close-packed array are
therefore optimally filtered for clusters of galaxies, with radio
sources constrained by the outlying telescopes.

\subsection{Field Selection}

The SZA survey fields were selected to lie far from the plane of the
Galaxy and to transit at high elevation at the OVRO site, minimizing
atmospheric noise while optimizing the imaging capabilities of the
array.  Fields were spaced equally in Right Ascension ($\alpha$) to
permit continuous observation. These constraints led to the selection
of four regions ranging in declination ($\delta$) from ${\rm
25\degree}$ to ${\rm 35\degree}$.  No consideration was
given to the location of known bright radio sources or clusters of
galaxies during field selection.

\begin{figure}[h!]
\begin{center}
\includegraphics[width=3.5in]{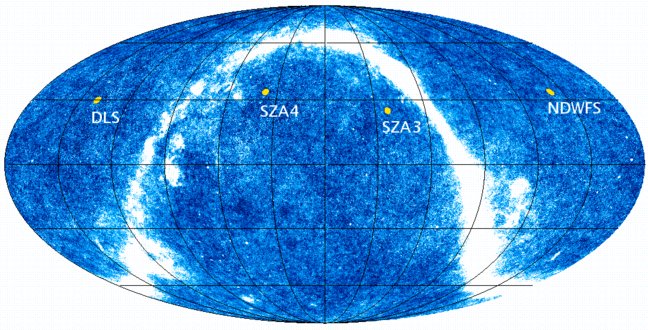}
\caption[IRAS dust map with overlay of the SZA field locations]{IRAS
dust map with overlay of the SZA field locations}.
\label{fig:fields}
\end{center}
\end{figure}

As the SZ effect itself provides no redshift information, the fields were
also selected to make maximum use of publicly available optical survey
data, from which photometric redshifts could be derived.  Two of the
fields were selected to overlap with regions for which optical data is
available, namely the Deep Lens Survey (DLS Field F2) \citep{dlsSurvey} and the
NOAO Deep Wide Field Survey (NDWFS) \citep{ndwfs}, also known as the Bootes
field.  Figure \ref{fig:fields} depicts the approximate locations of
the four fields on the IRAS ${\rm 100~\mu m}$ dust maps
\citep{iras}.

% REMOVE THE NEXT LINE
\vspace{0.1in}
\subsection{Data Collection}
\label{sec:trackStructure}

\begin{figure}[h!]
\begin{center}
\includegraphics[width=3.5in]{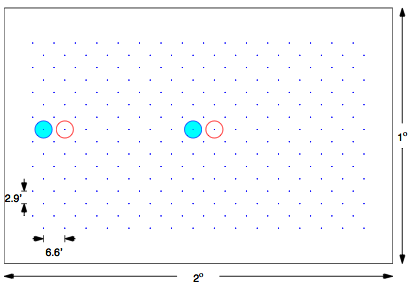}
\caption{Mosaic pointing locations for a given SZA survey field. The
fields are divided into 16 rows of 16 columns, with the pointings in
each row separated by 6.6\arcmin\ and each row offset from each other
by 2.9\arcmin.  This leads to each field being roughly 2 degrees by 1
degree in area.  In a single track the SZA observed four pointings
within a given row.  For example, pointings in the first and ninth,
followed by pointings in the second and tenth columns.}
\label{fig:trackStruct}
\end{center}
\end{figure}

Mapping large areas of sky with
an interferometer requires observing multiple discrete pointings within the
area to be covered, and combining the images from each pointing into a {\it mosaic}.  To perform such a combination of images, each of the four
fields is split into 16 rows of 16 pointings.  The pointings are
equally spaced by 6.6\arcmin \ along great circles in the $\alpha$
direction, and each row is equally spaced by 2.9\arcmin \ in the $\delta$
direction.  Subsequent rows are offset from one another so that that
the first pointing in each row is shifted by 3.3\arcmin \ in the $\alpha$
direction relative to the previous row.  This means that for a single
field we observe an area that spans roughly 2 degrees in the $\alpha$
direction and 1 degree in the $\delta$ direction  (see Figure
\ref{fig:trackStruct}).

For each of the survey fields, data were taken daily in 6 hour {\it
tracks}.  In a single track, we observed two staggered pairs of
pointings, within a single row.  These observations were performed in
a manner that permits ground subtraction from consecutive pointings in
a pair (although the ground contamination was found to be negligible
and was not subtracted as part of the analysis of this
paper).  Each track results in roughly 1 hour of observation on each
of the four pointings, with very nearly the same Fourier sampling for
pairs of pointings.  A second track is run at a later date, with the
order of the pairs reversed, to ensure that the Fourier sampling for
all four pointings is comparable.  In Figure \ref{fig:trackStruct} we
show the position of the pointings in each field, and indicate how the
pointings were observed in a given track.

For each set of four pointings, this sequence is repeated three times
over the span of roughly one year, so that each pointing is observed
in six tracks, translating to roughly 6 hours of observation per
pointing over the duration of the survey.

\subsection{Observations and Data Reduction}

%\begin{deluxetable}{lcrrccccccc}
\begin{deluxetable*}{lcrrccccccc}
\tabletypesize{\scriptsize}
\tablecolumns{8}
\setlength{\tabcolsep}{1mm}
\tablecaption{Survey Observations}
\tablehead{
\colhead{Field Name}& \multicolumn{2}{c}
{\underline{Field Center (J2000)}}& \multicolumn{2}{c}{\underline{Calibrators}} & \colhead{Dates} & \colhead{Integration} & \colhead{Rows} \\
\colhead{} & \colhead{$\alpha$}& \colhead{$\delta$} & \colhead{Bandpass (Jy)\tablenotemark{a}} & \colhead{Gain (Jy)\tablenotemark{a}} & \colhead{of Observations} & \colhead{Time (hrs)\
} &\colhead{Covered}}
\startdata
SZA4   & 02$^h$15$^m$38$^s$.3 &32$^{\circ}$08$^{\prime}$21$^{\prime\prime}$ & J2253+161 (11.6) & J0237+288 (2.9) & 07/11/2006 to 07/25/2007 & 687 & 7\\
DLS    & 09$^h$19$^m$40$^s$.0 &30$^{\circ}$01$^{\prime}$26$^{\prime\prime}$ & J0319+415 (11.0) & J0854+201 (5.4) & 11/18/2005 to 07/06/2007 & 1054 & 14\\
NDWFS  & 14$^h$30$^m$08$^s$.0 &35$^{\circ}$08$^{\prime}$34$^{\prime\prime}$ & J1229+020 (25.3) & J1331+305 (2.1) & 11/19/2005 to 07/23/2007 & 1000 & 14\\
SZA3   & 21$^h$30$^m$07$^s$.0 &25$^{\circ}$01$^{\prime}$26$^{\prime\prime}$ & J1642+398 (\phantom{2}5.5) & J2139+143 (1.4) & 11/13/2005 to 07/25/2007 & 1245 & 16\\
\enddata
\label{tab:obsTable}
\tablenotetext{a}{Fluxes obtained from 31~GHz SZA observations of sources on April 16, 2006. }
%\end{deluxetable}
\end{deluxetable*}

Images of the survey fields were produced by linear mosaicking of maps
from the individual pointings, as described in a companion paper,
\cite{muchovej2010}, hereafter \m10.  In particular, we stitch
together maps from the individual pointings, properly weighted by the
primary beam, to generate signal and noise maps of the fields.  We
further construct a {\it significance} (\snr) map by taking the ratio
of the signal and noise maps.  In Table \ref{tab:obsTable} we present
details of the mosaicked SZA survey.  The second and third columns
show the approximate center of each 16-row field.  We also present the
bandpass and gain calibrators in the next two columns, with their
fluxes as measured by the SZA (calibrated to observations of Mars, see
below).  In the fifth column we give the time range over which
observations were taken, with the caveat that observations were not
performed every day during that time span.  The penultimate column
lists the total integration time for data used in the analysis, and
the final column gives the number of rows observed in each field.  To
ensure uniform coverage of all fields, tracks were repeated when
deemed necessary due to poor weather or instrumental glitches.  Note
that the full 16 rows were not observed for all fields, due to
maintenance operations, instrumental characterization, and RFI
monitoring.  For the first 8 months of observations, the SZA4 field
was dedicated to CMB anisotropy measurements \citep{sharp2010}, and
data were collected in a manner incompatible with survey observations.
As a result, only 7 rows in the SZA4 field were completed to the full
survey depth.  We observed 6 more rows in that field, but not to the
same depth as the rest of the survey.  This results in a smaller
region of uniform sensitivity in SZA4, but the field is still usable
in our analysis.  In total, the data in the SZA cluster survey
correspond to 1493 tracks taken between November 13, 2005 and July 25,
2007.

Data for each track were calibrated using a suite of
MATLAB\footnote{The Mathworks,
Version 7.0.4 (R14), \tt{http://www.mathworks.com/products/matlab}}
routines, which constitute a complete pipeline for flagging,
calibrating, and reducing visibility data \citep[see][]{muchovej2007}.
Although the data were reduced exactly as described in that work,
survey data collection differed in a few key ways: whereas in targeted
observations we observed a source for 15 minutes before observing a
calibrator, in this work four distinct pointings were observed for
roughly 4 minutes each before observing a calibrator.  Also, system
temperature measurements were performed every eight minutes in survey
mode, as opposed to every 15 minutes in targeted observations.  The
absolute flux calibration is referenced to Mars, assuming the
\cite{Rudy1987} temperature model. Accounting for the uncertainty in the Mars model 
and in the transfer to our data, we assign a conservative uncertainty of 10\% to our flux
calibration.
Flagging of the data as described in \cite{muchovej2007} resulted in a
loss of roughly 23\%.  At the end of a single 6-hour
track, our on-source time per pointing was roughly 55 minutes, leading
to a noise level of approximately 1.5~mJy/beam in each pointing of the
short and long baseline maps.  Lastly we remove approximately 0.4\% of
the data with poor noise properties resulting from minor glitches in
the digital correlator.

\subsection{Mosaics}
\label{sec:mosaic}
Once data on all pointings in a given field are reduced, we construct
a linear mosaic of the field on a regular grid of 3.3\arcsec\
resolution.  This scale is much less than the requirement for Nyquist
sampling of the data, $\frac{1}{2D_{max}}$, where $D_{max}$ is the
longest baseline, and leads to a convenient number of pixels for the
use of FFTs in the following analysis.  The maps are composed of the
data across our 8~GHz of bandwidth, centered at a frequency of
30.938~GHz.  The attenuation due to the primary beam response 
for each pointing is corrected before mosaics are constructed. The
primary beam is calculated from the Fourier transform
of the aperture illumination of each telescope at the central
observing frequency, modeled as a truncated Gaussian with a central
obscuration corresponding to the secondary mirror.  Typical
synthesized beams for the short and long baseline maps for each of our
pointings have Gaussian FWHM of 2\arcmin\ and 45\arcsec~FWHM,
respectively.

Due to the overlap of neighboring pointings, the effective noise is
approximately uniform in the interior of the mosaics, but increases
significantly towards the edge of the mosaicked images.  We limit the
survey area by applying an edge cutoff in our mosaicked maps where the
effective noise is $> \noiseLim$~mJy/beam (corresponding roughly to
the one-third power point of the beam, given the noise in a single
pointing).

In Table \ref{tab:sensTable} we show the noise properties of the
observed fields.  We present the minimum and median noise (in
mJy/beam) for mosaic maps made with long baselines only, short
baselines only, and with the combination of the two.  The median noise
is calculated only in the region within which the noise is less than
the \noiseLim~mJy/beam cutoff.  The last column indicates the total
area covered in each field below the noise threshold.  That the
minimum and median pixel noise values are similar is an indication of
the uniformity of the coverage in the survey fields.  The SZA4 field
is not as uniform as the other fields, as only seven rows were
completed to full survey depth, while the remaining six rows were
observed for roughly half the time.

%\begin{deluxetable}{lccccccc}
\begin{deluxetable*}{lccccccc}
\tablewidth{0pt}
\tabletypesize{\scriptsize}
\tablecolumns{8}
\setlength{\tabcolsep}{2.3mm}
\tablecaption{Survey Sensitivity}
\tablehead{
\colhead{Field}& \multicolumn{2}{c}{\underline{Short Baselines}} & \multicolumn{2}{c}{\underline{Long Baselines}} & \multicolumn{2}{c}{\underline{All Baselines}} & \colhead{Area} \\
\colhead{Name} & \colhead{Minimum {\it rms}} & \colhead{Median {\it rms}} &  \colhead{Minimum {\it rms}} & \colhead{Median {\it rms}} &  \colhead{Minimum {\it rms}} & \colhead{Median \
{\it rms}} &  \colhead{Covered}\\
\colhead{} & \colhead{(mJy/beam)} & \colhead{(mJy/beam)} & \colhead{(mJy/beam)} & \colhead{(mJy/beam)} & \colhead{(mJy/beam)} & \colhead{(mJy/beam)} & \colhead{ (${\rm degree^2}$)}}
\startdata
SZA4       & 0.218 & 0.400 & 0.231 & 0.422 & 0.159 & 0.305 & 1.5 \\
DLS        & 0.201 & 0.237 & 0.200 & 0.250 & 0.142 & 0.173 & 1.5 \\
NDWFS      & 0.219 & 0.239 & 0.219 & 0.247 & 0.156 & 0.172 & 1.5 \\
SZA3       & 0.213 & 0.232 & 0.218 & 0.241 & 0.153 & 0.167 & 1.7 \\
\enddata
\label{tab:sensTable}
\end{deluxetable*}
%\end{deluxetable}

\subsection{Source Extraction}
\label{sec:sourceExtract}
Extra-galactic radio sources are a significant contaminant in
observations of the SZ effect,
particularly at centimeter wavelengths.  The source extraction
algorithm for the SZA survey consists of two stages, both of which use
5~GHz~VLA follow-up observations of our fields to assist in source
identification.  The first stage is an iterative fitting of bright
sources with fluxes at least \sigmaLim\ times the local map \rms.
This stage of fitting is described extensively in \m10\ and summarized in this section.
The second stage of source removal relies heavily on VLA follow-up
observations to remove dim sources.  A description of
the VLA follow-up observations, and their use in
determining parameters of the sources to be fit in the SZA data, can also be found in \m10.

Using the following algorithm we fit for 326 total sources: 239 in the
iterative stage of fitting (see \S\ref{sec:iterative}), and 87 in the
second stage of fitting (see \S\ref{sec:stage2}).  We note that of the
326 sources, 39 ($12\%$) were deemed to be extended in the follow-up
VLA data and fit as extended sources.  However, the fitted 31~GHz
major axes were all determined to be smaller than 22.5\arcsec (the
FWHM of the long baseline maps), implying they are mostly unresolved
by the SZA.  This allows us to approximate all sources detected in our
survey as unresolved by our long baselines in subsequent analysis (see
\S \ref{sec:completeness}).

\subsubsection{First Stage: Bright source removal}
\label{sec:iterative}
The first stage of the source extraction algorithm consists of
iterative removal of bright sources.  Source identification begins
in the image plane, with inspection of the combined (short and long
baseline) significance (\snr) maps for the brightest pixel with
significance greater than \sigmaLim.  Once we identify the location of
a source, we next determine whether the source is extended or
unresolved as seen by the VLA, and whether this candidate
is a single source, or a collection of nearby sources, using the
higher-resolution 5-GHz data obtained with the VLA.  Due to
the complex sidelobe structure of the synthesized beam, nearby sources
must be removed simultaneously from the interferometric data; we
therefore fit all sources within 45\arcsec\ of the primary
source location, roughly twice the synthesized beam width of the long
baseline maps.

Once we have identified all sources near the identified map peak
which are to be removed from the data, as well as their morphology
(compact/extended), we solve for source properties by fitting to the
multi-pointing visibility data.  For computational expediency, we
describe the sources as functions with analytic Fourier
transforms. Compact sources are treated as delta functions
characterized by a location, total intensity, and a spectral index
across our 8~GHz bandwidth. Extended sources are treated as elliptical
Gaussians, characterized by a location, integrated intensity, spectral
index across our band, ellipse eccentricity and angle of rotation.
Parameters for unresolved sources are fit only to the long-baseline
data, while those of extended sources are fit to all data.  The
best-fit models are removed from the Fourier data, and the mosaics are
regenerated.  This process is repeated iteratively until there are no
sources brighter than ${\rm\sigmaLim\sigma}$ in the significance maps.

In one case, the limited dynamic range of the SZA 
resulted in non-negligible residuals after source removal.  For this
(${\rm > 100~mJy}$) bright source, we remove it from our data, check
for the greatest residual level in the vicinity of the source, and
remove a section of our survey affected by the residuals which are not
within our noise properties.  This results in a hole (of
6.5\arcmin~diameter) in our coverage where the bright source is located.

\subsubsection{Second Stage: Faint source removal}
\label{sec:stage2}
The second stage of the fitting algorithm extracts dimmer sources,
relying heavily on our VLA 5~GHz follow-up.  We develop a source
catalog from follow-up data at 5~GHz and a spatial template of these
source.  Beginning with the catalog, we exclude those
sources already removed by the procedure of
\S\ref{sec:iterative} and examine the flux in the residual mosaics
at the positions of all remaining 5~GHz sources.  We consider 
any pixel whose \snr\ is greater than 3 (flux $\gtrsim0.5$~mJy) as a source candidate.  For these sources, we fit only for the flux; locations are fixed to the coordinates from the VLA, and 5/31~GHz spectral
indices are constrained at the lower frequency by the 
VLA fluxes.
For these dim sources, an unresolved source model ($\delta$-function)
fit to an extended source will result in residual structure in the map
that is within our noise properties, so we perform these fits only to
the long baseline data and remove the corresponding flux from the
short baseline map.

\section{Results}
\subsection{Survey Area}
\label{sec:surveyArea}
The principal data product of our survey is the number of clusters detected and the area observed as a
function of sensitivity.  Equipped with this information, a means to
translate from SZ flux to mass, a suitable mass function, and the
number of clusters detected in our fields, we can estimate $\sigma_8$,
the amplitude of mass density fluctuations on a scale of 8$h^{-1}$
Mpc.  Since the long baselines of the SZA are not sensitive to the
scales subtended by galaxy clusters, the following calculations
use only the source-subtracted, short-baseline data.

\begin{figure*}
\begin{center}
\includegraphics[width=7.5in]{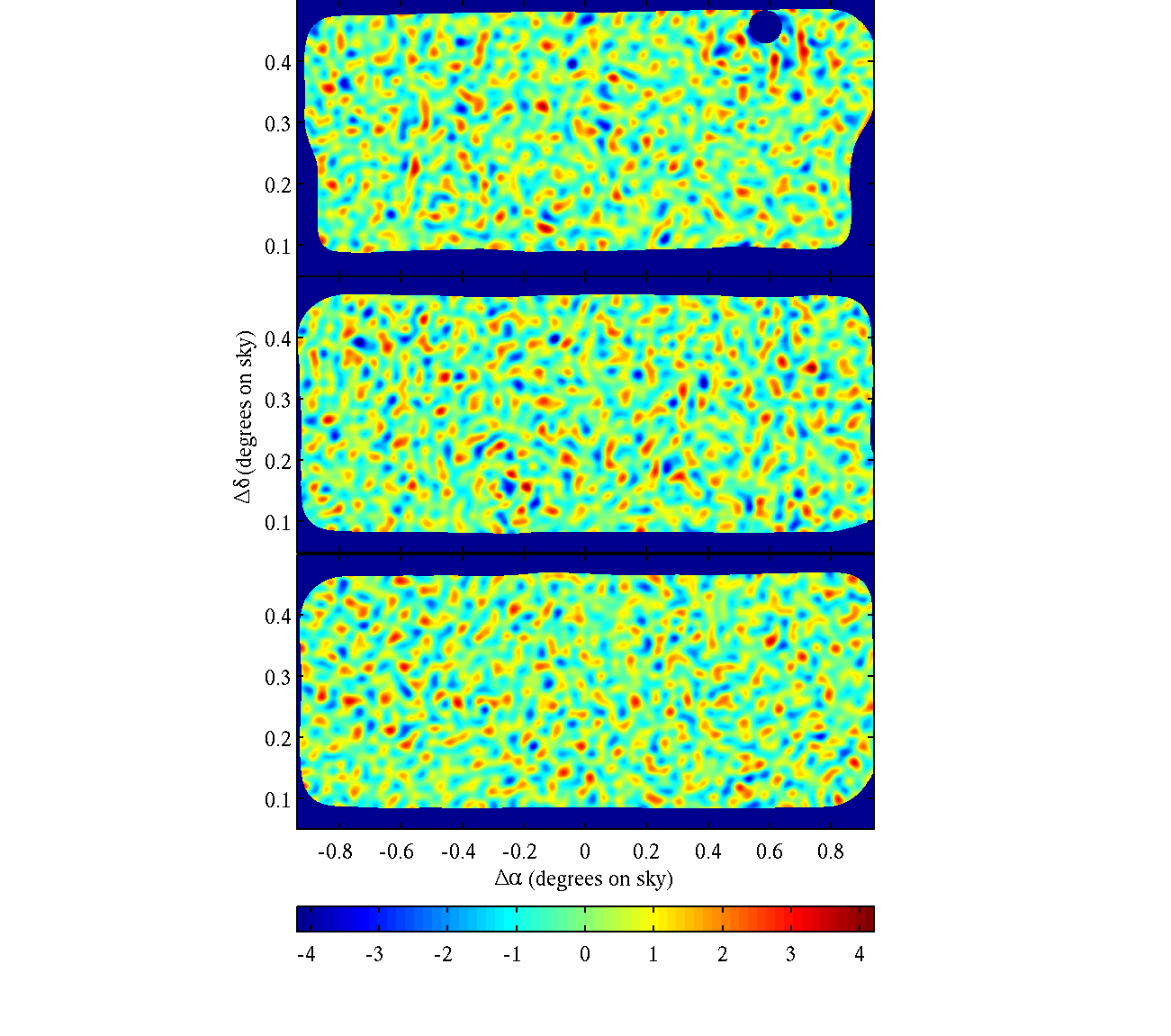}
\caption[Resulting significance maps]{Significance (\snr) maps for the SZA4,
DLS, and NOAO NDWFS fields after source removal and rescaling by the
noise.  Note the region excluded in the top map due to the limited dynamic range of the SZA.  We see no dark regions of significance greater than
4.3. }
\label{fig:maps}
\end{center}
%\end{figure}
\end{figure*}

We first calculate the sensitivity directly from our short-baseline
sky maps (which include diffuse emission), and compare this
to theoretical predictions.
One of our fields (the SZA3 field) shows excess noise
of 24\% over the expectation, and there is strong evidence for a spatial correlation
between the 31~GHz maps and ridges of dust observable in
the IRAS ${\rm 100\mu m}$ maps.  Lacking a suitable template for
removing this foreground from our data, we exclude the
field from our cosmological analysis.  This field is the subject of a companion paper (Leitch et al., in prep).
The rescaled, source-subtracted significance maps for the three remaining fields
are presented in Figure \ref{fig:maps}.

%\begin{figure}

\begin{figure}
\begin{center}
\includegraphics[width=3.0in]{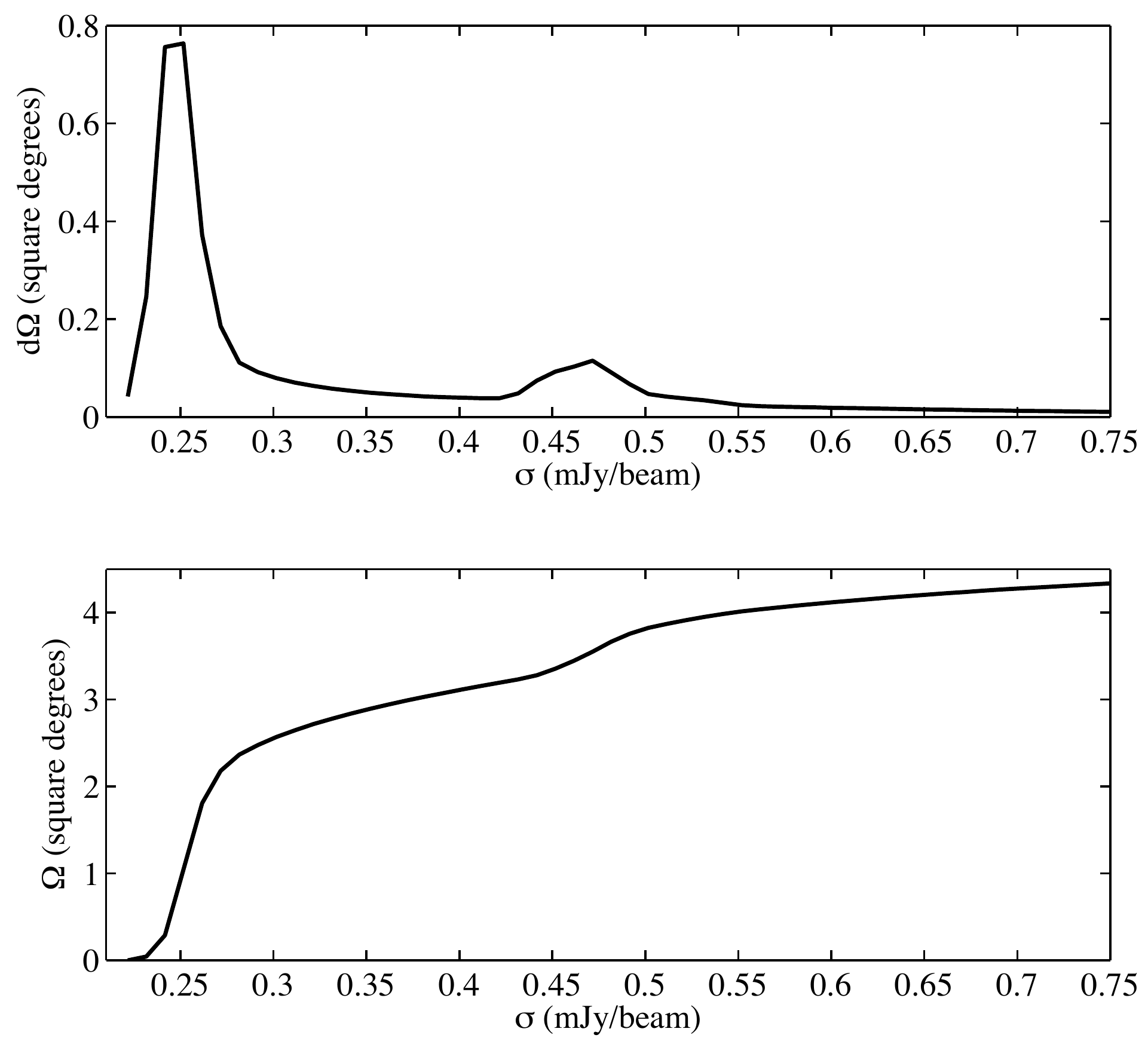}
\caption[Area as a function of Sensitivity]{Area as a function of
Sensitivity for the SZA survey.  In the top panel we show the
differential area as function of sensitivity ( d${\rm \sigma =
0.01~mJy/beam}$).  Notice the high bump near a sensitivity of
0.25~mJy/beam that encompasses most of our survey area.  The smaller
bump near a coverage of 0.47~mJy/beam is a result of the shallow
coverage on half of the SZA4 field.  In the bottom panel we present
the integrated area of the survey as a function of sensitivity.}
\label{fig:dOmegadSigma}
\end{center}
\end{figure}

\begin{deluxetable}{ccc}
\tabletypesize{\scriptsize}
\tablecolumns{2}
\tablewidth{0pt}
\tablecaption{SZA Area Coverage}
\tablehead{
\colhead{Noise Value} & \colhead{$\sum$ Area($<$noise)} \\
 \colhead{(mJy)} & \colhead{($\deg^2$)}
 }
 \startdata
0.22 & 0.000 \\ 
0.23 & 0.041 \\
0.24 & 0.287 \\
0.25 & 1.043 \\
0.26 & 1.807 \\
0.27 & 2.179 \\
0.28 & 2.365 \\
0.29 & 2.476 \\
0.30 & 2.568 \\
0.35 & 2.892 \\
0.40 & 3.116 \\
0.45 & 3.354 \\
0.50 & 3.823 \\
0.55 & 4.013 \\
0.60 & 4.120 \\
0.65 & 4.207 \\
0.70 & 4.278 \\
0.75 & 4.336 \\
\enddata
\label{tab:area}
\end{deluxetable}

We calculate numerically the differential survey area as a
function of \rms~ noise on intervals of d${\rm \sigma = 0.01~mJy/beam}$, shown in the top panel of Figure \ref{fig:dOmegadSigma}.  
Over most of the survey area, the noise lies between 0.25 and 0.3 mJy/beam.  In the bottom panel we present
the integral of the top panel to indicate the total survey area as
a function of sensitivity; the same data are presented in tabular format in Table \ref{tab:area}.

\subsection{Cluster Detection}
\label{sec:clusterDetection}
Once we have removed sources of emission and rescaled by the
noise, we search for clusters as decrements in the mosaicked maps.  We
consider a detection any pixel whose amplitude is at least \sigmaLim~
times the \rms~ noise level.  Note that the Fourier-space coverage of
the compact array was designed to match the typical cluster profile,
so that these mosaics have already been optimally filtered for cluster
detection.

No clusters were identified at $> 5\sigma$ significance, with the
largest decrement having a significance of 4.3$\sigma$.  This allows
us to place an upper limit on \sigE.

\section{Expected Number of Clusters}

\label{sec:outlineCalc}
The number of clusters of mass $M$ that we should detect, per unit noise and redshift interval, is
given by
{\scriptsize
\begin{eqnarray}
\label{eq:dNdMdsigma}
\nonumber {{dN}\over{dM\,d\sigma\,dz}}(M,\sigma, \sigma_8) &=& \\
p(D|M,z,\sigma) &\times& {{dN}\over{dM\,d\Omega\,dz}}(M,z,\sigma_8) \times {{d\Omega}\over{d\sigma}}(\sigma)
\end{eqnarray}}

\noindent where $p(D|M,z,\sigma)$ is the completeness, the probability of detecting a cluster of mass $M$ at redshift $z$ in the presence of noise
$\sigma$, ${\frac{dN}{dM d\Omega ~dz}(M,z,\sigma_8)}$ is the mass function, the predicted
density of clusters as a function of mass, redshift, and \sigE, and
$\frac{d\Omega}{d\sigma}(\sigma)$ is the survey area as a
function of noise level.

The total number of clusters we expect to detect for a given cosmology is then given by the
integral of Equation \ref{eq:dNdMdsigma} over mass, redshift and map noise:
{\scriptsize
\begin{equation}
N(\sigma_8) = \int{dz}\int{d\sigma}\int{{{dN}\over{dM\ d\sigma ~dz}}(M,\sigma,\sigma_8)\,dM.}
\label{eq:N_int}
\end{equation}}
Here we assume the concordance cosmological parameters from WMAP 7-year results \citep{larson2010}.

\subsection{The Mass Function}
\label{sec:mass_function}
We assume the mass function of dark matter halos derived from
cosmological simulations of flat ${\rm \Lambda CDM}$ cosmology.  We
adopt the fitting formula of \cite{tinker2008}, and assume a redshift
evolution for the over-density parameter given by the growth factor of
\cite{viana1996}.

\subsection{Completeness}
\label{sec:completeness}
\label{sec:compSims}
As indicated in Equation \ref{eq:y}, the fundamental SZ observable is the
compton-$y$ parameter.  We can relate this observable to the
cluster mass either through observation or simulation, both of which
are subject to significant uncertainties.  SZ-effect and X-ray
observations can be used to determine scaling relations between
compton-$y$ and cluster mass \citep[e.g.,][]{bonamente2008}, but these
observations typically comprise a small number of massive clusters
spanning a limited redshift range, and often relate quantities
determined at overdensity radii not directly comparable to those
sampled by a particular experiment. By contrast, simulations can be
accurately compared to experimental details, but the correspondence
between the SZ observable and cluster mass is highly dependent on the
accuracy of the simulated gas thermodynamics \citep{kravtsov2005}.

We use the simulation of \cite{shaw2009} (hereafter \shaw09) to
translate between compton-$y$ and cluster mass.  This simulation
combines an N-body `lightcone' simulation with a semi-analytic model
for the cluster gas, with a significant amount of heating from
feedback processes (i.e., AGN feedback, star formation).  Gas
parameters in \shaw09\ have also been adjusted to match X-ray
observations of low-redshift clusters.  We select clusters from 40
compton-$y$ maps, each 5 degrees on a side, with ${\rm M_{200}}$
masses ranging from ${\rm 2.5\times 10^{13}~M_{\odot}}$ to ${\rm 3.0
\times 10^{15}~M_{\odot}}$, where ${\rm M_{200}}$ is the mass enclosed
within a radius corresponding to an overdensity of 200 times the mean
density of the universe; it is this mass that we refer to in all
subsequent discussion.

To calculate the probability that the SZA would detect a cluster of a
given mass, mock observations of the y-maps are performed in survey
mode, in a multi-pointing mosaic scheme.  For computational
expediency, we do not simulate observations of large areas of
simulated sky, but select individual clusters from the catalog and
calculate what fraction of them would have been detected.  Each
simulated cluster is placed at the center of a small field,
0.47$\degree$ on a side.  On this image we
overlay a grid of pointings in a 2-3-2 hex pattern, reflecting the
spacings in our survey.  The cluster image is weighted by the primary
beam (centered at each pointing), and the result is
Fourier-transformed and re-sampled onto a $uv$-grid that reflects the
actual coverage of the survey data.  To the visibilities in each
pointing we add Gaussian noise, with weights chosen such that the
resulting noise in the image plane mosaic is uniform in a region
within a 4\arcmin\ radius of the cluster.  To quantify the cluster
detectability as a function of noise, we generate maps with \rms\
noise between $0.2 - 0.65$~mJy/beam, in steps of $0.05$~mJy/beam.

To simulate the effect of compact source contamination, we add
unresolved sources brighter than 0.05~mJy to the data according to the
\m10\ distribution for regions further than 0\arcmin.5 from the
cluster center.  To account for cluster-source correlations we add
sources brighter than 0.01~mJy according to the distribution from
\cite{coble2007} for regions near the cluster center.  To
mimic our source-extraction algorithm, we generate a 5~GHz flux for
each source from the spectral index distribution of \m10, and create a
5-GHz source catalog of all sources brighter than 0.35~mJy (the
detection threshold for our VLA 5-GHz source catalog).  Sources are
extracted from the simulated maps exactly as described in \S
\ref{sec:sourceExtract}, with the exception that we remove a perfect
model from the simulated data rather than a fitted flux and spectral
index, in the interest of computational speed.  We have verified
that this does not systematically bias our results by comparing the
fitted flux of sources in these simulations with their original
values, i.e., the presence of noise introduces a random error into the
fit flux of the source, yet over the many realizations performed we
recover the true flux of the source.

Once the identified sources are removed from the mock observation, we search for
the peak decrement in the short-baseline mosaics, within the region of
uniform noise in our mock observation (namely 4\arcmin).  We identify
clusters in the simulated data as described in \S \ref{sec:clusterDetection}.

To capture the inherent flux distribution for clusters of similar
mass, as well as the redshift evolution of clusters, clusters were
observed in nine different mass ranges ($M_{\rm min} = {\rm 1.5
\times 10^{14}~M_{\odot}}$, $\Delta M = {\rm 10^{14}~M_{\odot}}$)
and eight redshift bins ($z_{\rm min} = 0.1$, $\Delta z = 0.2$).  
The number of clusters observed in each mass range and
redshift bin varied from 2 to 200; some bins contained no
clusters, namely the high-mass, high-redshift bins.
We further select only clusters which are not within 0.23$\degree$ of
the edge of the simulated maps, and which are not within 4 arcminutes of clusters of
greater or comparable masses.  The first cut ensures we do not
introduce artifacts associated with field edges, and the second
ensures that our observations are of clusters whose masses are
well-defined, i.e., whose simulated masses are not corrupted by the
presence of secondary clusters.  For each mass and redshift bin, we
iterate over 200 distinct realizations of clusters, noise and compact
sources.

To calculate the resulting completeness, we determine the fraction of
time the clusters in a given bin were detected at significance
greater than ${\rm 5\sigma}$.

\subsection{Survey Mass Limit}

\begin{figure}
\begin{center}
\includegraphics[width=3.5in]{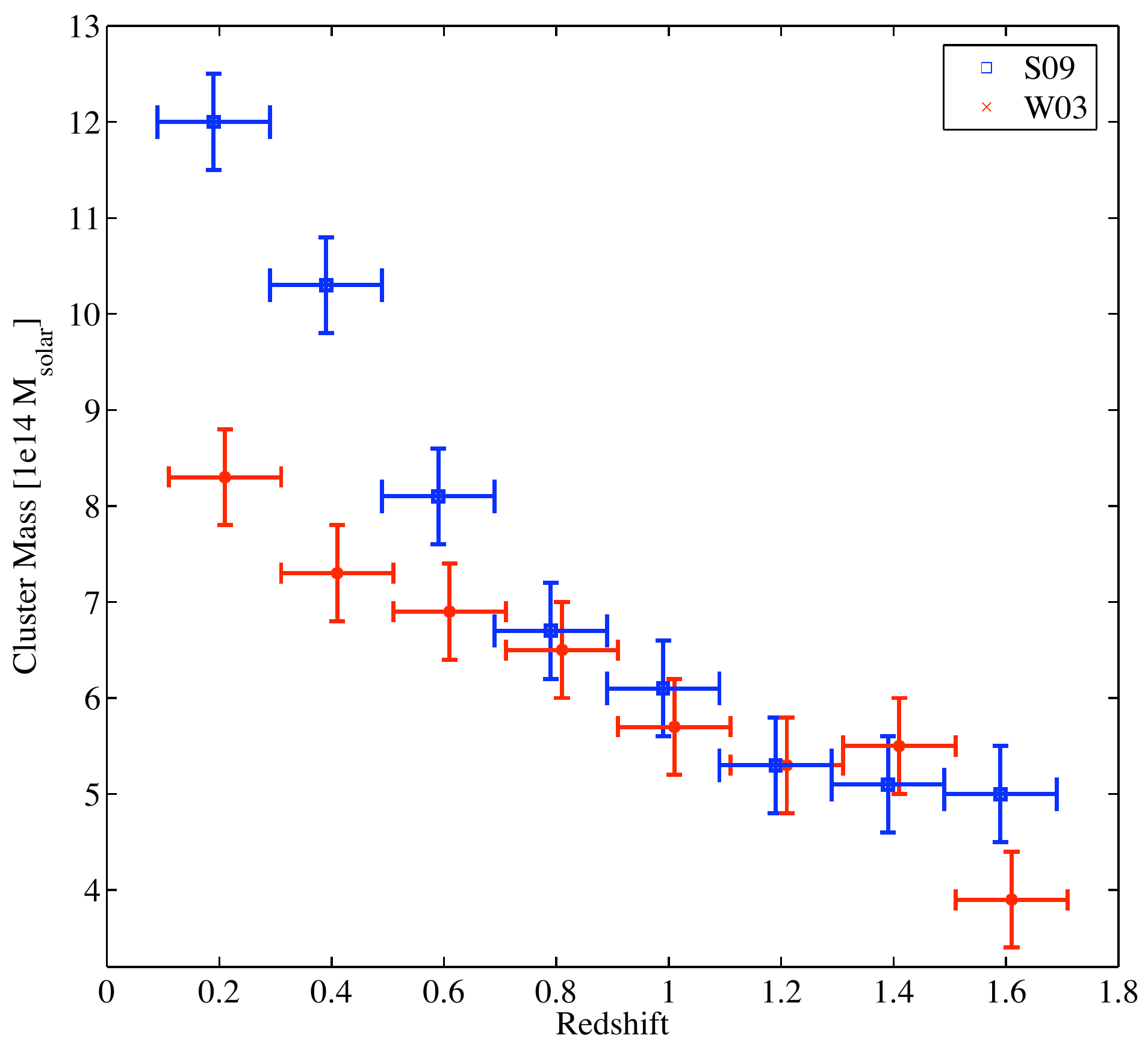}
\caption[SZA Mass Limit]{The mass limit of the SZA survey calculating
using the area as a function of noise, and cluster detectability from
simulations of \shaw09 (square points) and \white03 (See \S
\ref{sec:errors}).  The points correspond to the ${\rm M_{200}}$ mass
of a cluster which would have been detected at least 50\% of the time, 
and are offset from each other in redshift for ease of presentation.
The differences between values are a result of the \white03
simulations not including AGN feedback or ``pre-heating'' in
calculating the compton-$y$ parameter.}
\label{fig:massLim}
\end{center}
\end{figure}

Equipped with the survey completeness (\S \ref{sec:completeness}) and
the area as a function of noise (\S \ref{sec:surveyArea}), we can
calculate the area as a function of mass and redshift.
As the SZ effect depends most directly on cluster mass, this can be
used to calculate the redshift-dependent mass limit of our survey,
which we present in Figure \ref{fig:massLim} for the \shaw09
simulations.
For each redshift bin, we present the ${\rm M_{200}}$ mass above which
our survey is 50\% complete.  As expected, the mass limit decreases
towards higher redshifts, as clusters become more compact and
therefore more easily detectable by the SZA.  

In Figure
\ref{fig:massLim}, we also present the limiting mass calculated from a
different simulation, \cite{white2003}, to demonstrate the effect of
gas physics in our completeness calculations.  The differences between
the simulations and their effect on our results are discussed in
detail in \S \ref{sec:errors}.

\begin{figure}
\begin{center}
\includegraphics[width=3.5in]{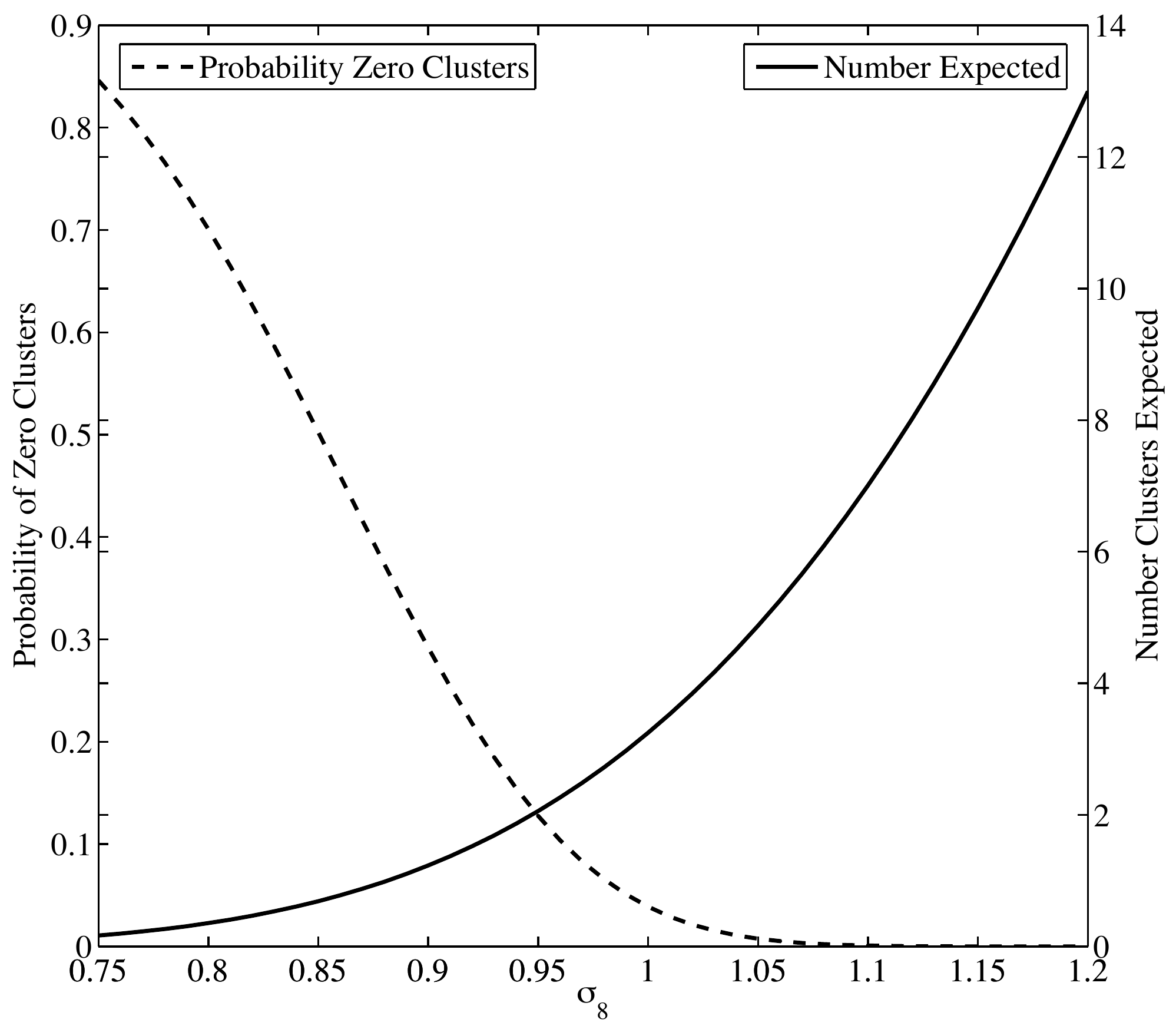}
\caption[Expected Number of Clusters Seen by the SZA Survey]{Expected
number of clusters (and the probability of non-detection of clusters)
in the 4.4 square degree SZA survey, as a function of \sigE, using
WMAP cosmology for all other parameters and cluster detectability from
\shaw09 simulations.  The solid line represents the expected number of
clusters, with the corresponding scale presented on the right axis.
The dashed line is the probability of detecting zero clusters, with
the corresponding scale on the left axis.}
\label{fig:numClusters}
\end{center}
\end{figure}

\section{Constraint on \sigE}
\label{sec:bayesianCalc}

To calculate the value of \sigE\ which is most consistent with the SZA
survey, we address the question: what is the probability of a \sigE\ value given
the number of observed clusters, $P(\sigma_8|N)$?  A simple invocation of Bayes' theorem yields:
\begin{equation}
\label{eq:bayesSig8}
P(\sigma_8|N) \propto P(N|\sigma_8) P(\sigma_8)
\end{equation}
where $P(N|\sigma_8)$ is the probability of $N$ detections given a value of $\sigma_8$ and $P(\sigma_8)$ is the prior on \sigE, which we take to be uniform.

The number of clusters detected is a Poisson process, with a
well-defined distribution given by 
\begin{equation}
\label{eq:poisson}
P(k|\lambda) = \frac{e^{-\lambda} \lambda^k}{k!}
\end{equation}
where $P(k|\lambda)$ is the probability that a process with expectation $\lambda$
occurs $k$ times. 
As we detect no cluster candidates in the SZA survey (at a significance greater than $5\sigma$), we have $k$=0, whence
\begin{equation}
\label{eq:probFinal}
P(\sigma_8|N) = e^{-N(\sigma_8)}
\end{equation}
where $N(\sigma_8)$ is the result from Equation \ref{eq:N_int}, shown
pictorially in Figure \ref{fig:numClusters}.  
Note that in determining $N(\sigma_8)$, we adopt concordance cosmological parameters from WMAP 7-year results \citep{larson2010}, and
calculate for completeness determined from the 
\shaw09 simulations.
Figure~\ref{fig:numClusters} (via Equation~\ref{eq:probFinal}) can be
used to compare the relative likelihood of different values of \sigE.
For example, we expect a non-detection in the SZA survey $5\%$ of the
time if $\msigE = 0.97$.

\begin{figure}
\begin{center}
\includegraphics[width=3.5in]{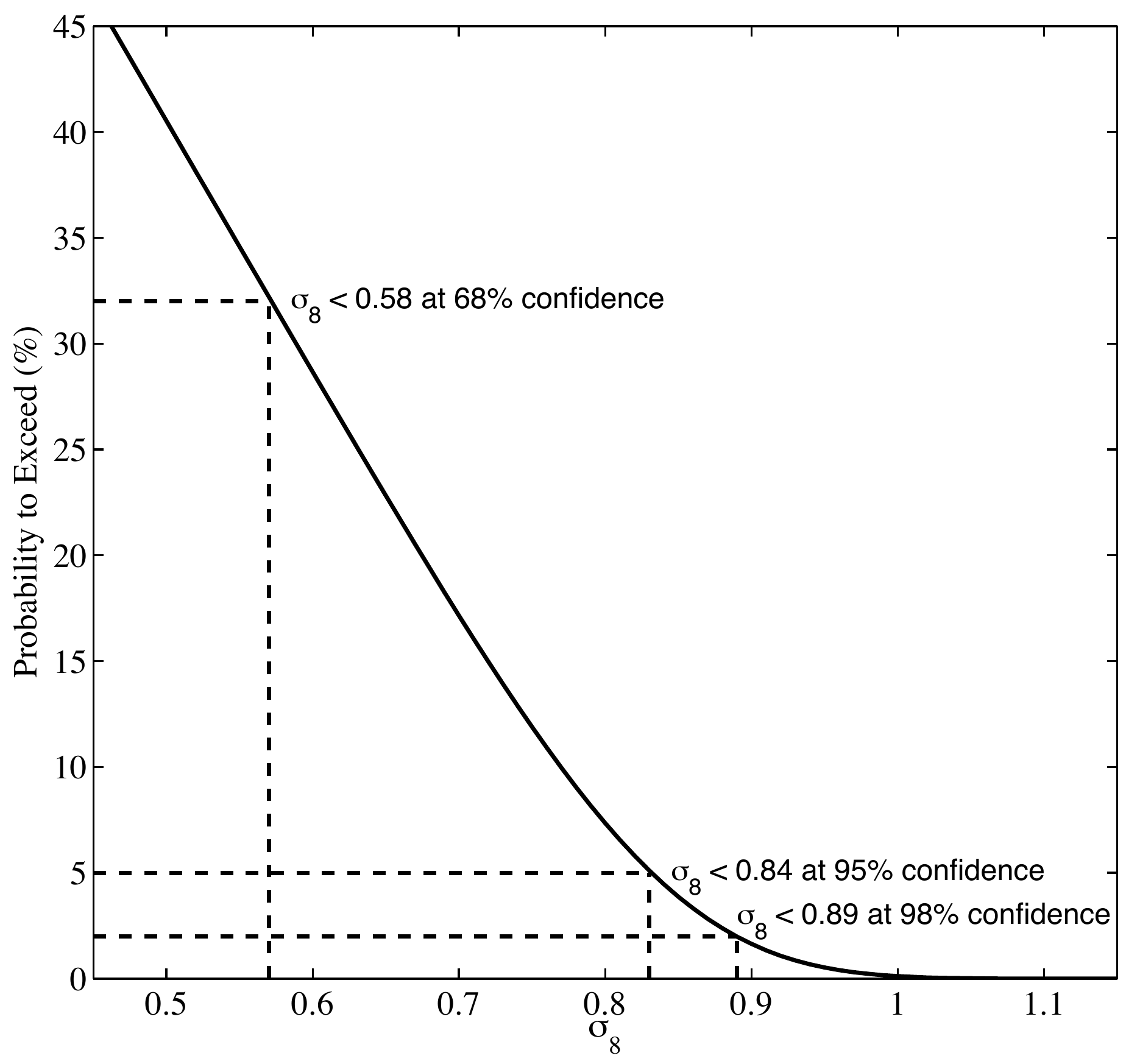}
\caption[Likelihood of a value of \sigE]{Plot of the probability of
that \sigE\ exceeds a given value for completeness derived using the
\shaw09 simulations, with the 1-, 2-, and 3-$\sigma$ confidence limits
presented.}
\label{fig:probSig8}
\end{center}
\end{figure}

We integrate Equation~\ref{eq:probFinal} to generate the probability,
given our data, that \sigE\ exceeds a given value; the result is
presented in Figure~\ref{fig:probSig8}.  We see that for the \shaw09
simulation, \sigE\ values of greater than 0.84 are ruled out at 95\%
confidence. 

In calculating this limit, we have made the conventional assumption of
a ``non-informative'' prior (i.e., uniform for $\msigE\ > 0$).  While
we consider this approach conservative, in that it captures what the
SZA data can say about the value of \sigE\ under a minimal set of
prior assumptions, note that it will also lead to the tightest
possible (for uniform priors) constraint on \sigE.  If for example,
the prior were truncated to $\msigE\ > 0.5$, our $95\%$ limit would
shift to $\msigE < 0.87$.  Sensitivity to the prior is typical for
data sets with unequal power to discriminate among allowable
hypotheses; in this case, the small area of the SZA survey can
strongly rule out high values of \sigE, but provides little or no
power to discriminate against low values of \sigE, to which our prior
nonetheless assigns equal weight.

\section{Systematics}
\label{sec:errors}

The conversion from cluster mass to SZ observable is potentially the
largest systematic uncertainty in cosmological analyses of SZ surveys.
Because constraints on \sigE\ are determined by comparison with
simulations, they are necessarily sensitive to the assumptions that
underlie each simulation's modeling of the ICM.
As a rough estimate of the importance of model assumptions to our
limit on \sigE, we repeat our analysis using the simulation of
\cite{white2003} (hereafter \white03), a high-resolution N-body
simulation following the semi-analytic method of \cite{schulz2003}.
The salient difference between the \shaw09 and \white03 simulations is
that \shaw09\ incorporate AGN and supernovae feedback while \white03\
ignore cold gas and star formation.  As a result, clusters in the
\white03\ simulation tend to be more compact (and therefore more
detectable by the SZA), with higher central SZ-decrements than
equivalent clusters (by mass) in \shaw09.  This leads to a higher
completeness for \white03\ clusters of a given mass, a lower mass
limit (See Figure \ref{fig:massLim}), and a stronger constraint on
\sigE\ (a 95\% confidence limit that is lower by 0.04).  We choose to
present the limit derived from \shaw09\ in \S\ref{sec:bayesianCalc}
not only because it is the more conservative of the two, but also
because the \shaw09 simulation represents a more
realistic model of ICM physics; we nevertheless caution that
significant uncertainties in the modeling of cluster gas physics
remain.

In the completeness calculation described in \S\ref{sec:completeness},
we model the correlation between clusters and radio sources by using
the \cite{coble2007} source distribution for the inner regions of
clusters, and the \m10\ distribution for field sources.  However the
\cite{coble2007} distribution was determined from observations of the
most massive clusters, and can bias our completeness low if sources are
over-represented in these objects relative to the lower-mass clusters
to which the SZA is sensitive.  To bracket the magnitude of this
effect, we repeat the completeness calculations described in \S
\ref{sec:completeness} without using the \cite{coble2007}
over-densities, i.e., using the source counts obtained only from field
sources presented in \m10.  Neglecting the higher density of sources
towards galaxy clusters increases our completeness on the order
of 12\%, which due to the steepness of the dependence of cluster
counts on \sigE\, would lower our limit by 0.04.

In \S \ref{sec:sourceExtract} we mentioned that for one bright source,
the limited dynamic range of the SZA introduces a hole in our coverage
at the source location.  If the correlation between clusters and radio
sources had not properly been taken into account in calculating our
completeness, this would bias our completeness low if bright sources
are preferentially associated with galaxy clusters.  However, we
simulate both the correlation and the corresponding reduction in
completeness entailed by missing clusters associated with sources that
we cannot subtract; cutting such sources out of our maps therefore
corresponds to a simple reduction in the survey area, and should have
no effect on our limit.

Although we see no evidence of sources which are resolved by the
long-baseline data, as discussed in \S \ref{sec:sourceExtract}, we fit
the 12\% of sources deemed to be extended at 5~GHz to a combination of
the long and short-baseline data (whereas sources deemed {\it a
priori} to be unresolved were fit only to the long-baseline maps).  If
any of these sources happened to be associated with a cluster, the
cluster decrement present in the short-baseline data would reduce the
fitted flux of the source, resulting in residual source flux in the
map that would bias against detecting the cluster.  To quantify this
effect, we re-calculate the completeness with a slight modification in
12\% of the realizations, namely that we do not remove the true flux
of the source, but instead remove the map flux at the source location
in the combination of short and long baseline maps.  As the majority
of sources are not associated with clusters, this test can be thought
of an upper limit to any potential reduction in our completeness.
Properly accounting for this effect would raise our limit on \sigE\ by
at most 0.01.

An additional source of uncertainty unrelated to our analysis methods
is the clustering of large-scale structure.  We know that galaxy
clusters are not evenly distributed throughout space, but form
preferentially along filaments.  As a result, the number of clusters
seen in small fields such as those observed by the SZA do not follow a
Poisson distribution.  In particular, since most lines of sight will
sample the voids between filaments, this leads to an increased
probability of detecting fewer massive halos than the Poisson average.
To quantify the impact on our results, we
begin with the 100 square degree \white03 simulations.  We select
regions similar in shape, size and noise properties to each of our
fields from the simulated maps, and fold in the calculated
completeness to estimate the number of clusters we should detect.  We
do this for $> 10^4$ realizations of each of the three SZA fields,
varying the location and orientation of the field over the sky-map, to
generate a distribution of expected numbers of clusters, and compare
this to the Poisson prediction for the equivalent field size and input
\sigE.  The result is a 10\% increase in the probability of a
non-detection for an input \sigE\ of 0.9 over the Poisson prediction,
leading to an underestimate of the true \sigE.  Accounting for this
effect, assuming similar clustering over a range of \sigE, would
raise our limit by at most 0.04.

Lastly, as described in M10, the calibration of the SZA data is tied
to the modeled flux of Mars, which we estimate is accurate to
$\lesssim10\%$.  A shift in the flux of Mars corresponds to a simple
rescaling of the map noise, and consequently a shift in the
completeness.  As noted above, a change in the completeness of
$\lesssim10\%$ corresponds to a shift in our limit on \sigE\ (in either
direction) of $\lesssim 0.04$.

\section{Discussion}

From 2005 to 2007, the SZA performed a ~6.1 square degree survey for
clusters of galaxies via their SZ effect at 31~GHz.  In one of the
fields there is evidence for large-scale dust-correlated emission;
this field was excluded from the analysis presented here.  Of the
remaining 4.4 square degrees of the survey suitable for cosmological
analysis, we estimate that the survey is $50\%$ complete to a mass of
${\rm M_{200 \overline{\rho}} \sim 6\times 10^{14}~M_{\odot}}$,
averaged over redshift.  By comparison with simulations, we place an
upper limit on the value of \sigE\ of ${\rm 0.84~(+ 0.07)}$ at 95\%
confidence, where the uncertainty reflects calibration and systematic
uncertainties discussed in \S \ref{sec:errors}, excluding the error
associated with the simulated cluster gas physics.  Although this last
uncertainty is potentially the dominant one, to properly quantify it
requires a calculation of the completeness for a wide range of
simulations with realistic gas models, which is beyond the scope of
this paper.

Our limit on \sigE\ is consistent with recent results from SZ surveys performed
over larger areas of sky, such as with the South Pole Telescope
\citep{vanderlinde2010}, and with determinations of \sigE\ from
gravitational lensing and X-ray cluster surveys \citep{smith2003,
allen2003}.  In addition, it is consistent with determinations of
\sigE\ from CMB anisotropy measurements, namely those of WMAP
\citep{dunkley2009}, the South Pole Telescope \citep{lueker2010}, and
the Atacama Cosmology Telescope \citep{act2010}.  Our constraint is
also in agreement with that of \cite{sharp2010}, based on CMB
anisotropy measurements with the SZA itself.  Although
the data were collected with the same instrument, we note that both the data sets and
the analyses of this paper and of \cite{sharp2010} are completely
independent.

\acknowledgements We thank John Cartwright, Ben Reddall and Marcus
Runyan for their significant contributions to the construction and
commissioning of the SZA instrument.  We thank the staff of the Owens
Valley Radio Observatory and CARMA for their outstanding support.  We
thank Bryan Butler and Mark Gurwell for providing the Mars model to
which the SZA data are calibrated.  We also thank Laurie Shaw and Martin
White for providing us with the simulations used in this work.  We
gratefully acknowledge the James S.\ McDonnell Foundation, the
National Science Foundation and the University of Chicago for funding
to construct the SZA.  The operation of the SZA is supported by NSF
Division of Astronomical Sciences through grant AST-0604982. Partial
support is provided by NSF Physics Frontier Center grant PHY-0114422
to the Kavli Institute of Cosmological Physics at the University of
Chicago, and by NSF grants AST-0507545 and AST-05-07161 to Columbia
University.  AM acknowledges support from a Sloan Fellowship, and SM
from an NSF Astronomy and Astrophysics Fellowship, and CG, SM, and MS
from NSF Graduate Research Fellowships.

{\it Facilities:} {SZA}, {VLA}

\bibliographystyle{apj}

\bibliography{sza}

\end{document}